\documentclass[12pt,a4paper]{article}
\usepackage{amsmath}
\usepackage{amssymb}
\usepackage{graphicx,epsfig}
\usepackage[mathscr]{eucal}
\usepackage[numbers,sort&compress]{natbib}
\usepackage[english]{babel}
\usepackage{tabularx}
\usepackage{float}
\usepackage{slashed}

\usepackage[colorlinks=true,allcolors={black}]{hyperref}

\usepackage{scalefnt,ulem,pstricks}

\usepackage{setspace}
\usepackage{xspace}

\usepackage{multirow}

\newcommand\as{\alpha_{\mathrm{S}}}

\def\to{\rightarrow}
\def\nn{\nonumber}

\newcommand{\yttwth}{\tilde y_{23}}
\newcommand{\ktFSR}{k_T^{\rm FSR}}
\newcommand{\xT}{x_T^{\rm FSR}}

\usepackage{scalefnt,pstricks}
\usepackage{cancel}

\usepackage{array}
\newcolumntype{L}[1]{>{\raggedright\let\newline\\\arraybackslash\hspace{0pt}}m{#1}}
\newcolumntype{C}[1]{>{\centering\let\newline\\\arraybackslash\hspace{0pt}}m{#1}}
\newcolumntype{R}[1]{>{\raggedleft\let\newline\\\arraybackslash\hspace{0pt}}m{#1}}

\usepackage{lipsum}

\allowdisplaybreaks

\begin{document}
\begin{titlepage}
\begin{flushright}
  ZU-TH 74/23 \\
  CERN-TH-2023-209\\
\end{flushright}

\renewcommand{\thefootnote}{\fnsymbol{footnote}}
\vspace*{0.5cm}

  \vspace{5mm}

\begin{center}
  {\Large \bf Subleading power corrections\\[0.2cm]for event shape variables in $e^+ e^-$ annihilation}
\end{center}

\par \vspace{2mm}
\begin{center}
  {\bf Luca Buonocore${}^{(a,b)}$, Massimiliano Grazzini${}^{(a)}$,\\[0.2cm] Flavio Guadagni${}^{(a)}$,}
and    
{\bf Luca Rottoli${}^{(a)}$}

\vspace{5mm}

${}^{(a)}$Physik Institut, Universit\"at Z\"urich, CH-8057 Z\"urich, Switzerland\\[0.2cm]
${}^{(b)}$CERN, Theoretical Physics Department, CH-1211 Geneve 23, Switzerland

\vspace{5mm}

\end{center}

\par \vspace{2mm}
\begin{center} {\large \bf Abstract}

\end{center}
\begin{quote}
\pretolerance 10000

We consider subleading power corrections to event shape variables in $e^+e^-$
collisions at the first order in the QCD coupling $\as$. We start from the
jettiness variable $\tau_2$ and the $y_{23}$ resolution variable for the
$k_T$ jet clustering algorithm and we analytically compute the corresponding cumulative cross
section. We investigate the origin of the different power suppressed
contributions in the two-jet limit and trace it back to
their different coverage of the phase space.
We extend our analysis to the case of thrust and of the $C$-parameter, and we finally discuss a class of observables
that depend on a continuous parameter giving different weight to central and forward emissions
and we evaluate the corresponding subleading power corrections.

\end{quote}

\vspace*{\fill}
\begin{flushleft}
November 2023
\end{flushleft}
\end{titlepage}

\section{Introduction}
\label{sec:intro}

Event shapes and jet rates have been extensively studied in $e^+e^-$ collisions (see e.g. Ref.~\cite{Dasgupta:2003iq} and references therein).
The former measure geometrical properties of the final-state hadronic energy flow, while the latter 
allow us to {\it count} the number of jets, thereby providing access to the underlying partonic structure of the hadronic event. 
Since jet rates always depend on a resolution parameter, they can themselves be used to define event shape variables.

The value of a given event shape encodes in a continuous fashion, for example, the transition from pencil-like two-jet events
to planar three-jet events or to events with a spherical distribution of hadron momenta. For this reason,
event shapes were already widely used in early studies of strong interactions.
Being infrared (IR) safe by construction, event shapes and jet rates can be computed order by order in perturbation theory, and can in turn be used to measure the QCD coupling $\as$.
More generally, these observables are also relevant in studies of the interplay between perturbative and non-perturbative QCD.

In this paper we focus on event shape variables that are non-zero in three-jet configurations.
We generically denote an event shape variable (that we assume to be properly normalised to make it dimensionless) as $r$,
such that the two-jet limit corresponds to $r\to 0$. The differential cross section in this limit
receives large logarithmic contributions that need to be resummed to all orders.
Such resummation has been extensively studied \cite{Catani:1991kz,Catani:1991bd,Catani:1991pm,Catani:1992ua,Catani:1992jc,Dissertori:1995np,Catani:1998sf,Dokshitzer:1998kz,Becher:2008cf,Abbate:2010xh,Monni:2011gb,Mateu:2013gya,Hoang:2014wka,Banfi:2016zlc} at leading power, and observable-independent formulations of the resummation program do exist \cite{Bonciani:2003nt,Banfi:2003je,Banfi:2004yd,Banfi:2014sua,Bauer:2019bsp}.

The {\it next-to-leading power} contributions in the $r\to 0$ limit have received less attention, and only recently they have started to be systematically investigated \cite{Moult:2016fqy,Boughezal:2016zws,Moult:2018jjd,Moult:2019mog,Moult:2019uhz,Beneke:2020ibj,Beneke:2022obx,Agarwal:2023fdk}.
Besides helping us to improve our understanding of perturbative QCD, the study of power suppressed contributions is important when the observable is used as resolution variable to set up
higher order computations with {\it non-local subtraction} or {\it slicing} schemes \cite{Giele:1991vf,Catani:2007vq,Stewart:2010tn,Abreu:2022zgo,Buonocore:2022mle}.

In this paper we study subleading power corrections for several different event shape variables.
We start from the jettiness $\tau_2$ \cite{Stewart:2010tn}
and $y_{23}$ resolution variable for the $k_T$ jet clustering algorithm \cite{Catani:1991hj}.
We compute the necessary ingredients to use them as slicing variables to evaluate generic $e^+e^-\to 2$ jet observables at next-to-leading order (NLO).
We show that while the linear power corrections for jettiness are logarithmically-enhanced, those for $y_{23}$ are not. We also contrast the behavior of $\tau_2$ and $y_{23}$ with
that of a toy variable $\ktFSR$, which can be defined at NLO as the transverse momentum of the gluon with respect to the quark-antiquark pair.
Then, we analytically compute the
cumulative cross section for these observables, and discuss the origin of the different behavior of power corrections,
which is traced back to the different way in which the phase space is covered by these variables.
We then move to the thrust \cite{Farhi:1977sg} and $C$-parameter \cite{Parisi:1978eg,Fox:1978vu,Donoghue:1979vi},
evaluating the corresponding power corrections and discussing their origin.
We finally consider a variable $r_b$ depending on a continuous parameter $b$ that gives different weight to central and forward emissions along the relevant collinear direction, and we compute the ensuing subleading power corrections.

The paper is organised as follows. In Sect.~\ref{sec:notation} we introduce our notation and discuss the implementation of $\ktFSR$, $\tau_2$ and $y_{23}$ as resolution variables.
In Sect.~\ref{sec:calculation} we carry out our analytical study. We first compute the cumulative cross section for $\ktFSR$ (Sect.~\ref{subsec:xt}), $\tau_2$ (Sect.~\ref{subsec:tau2}) and $y_{23}$ (Sect.~\ref{subsec:y23}) and in Sect.~\ref{subsec:comparison-tau2-y23} we discuss their physical differences.
Then in Sec.~\ref{subsec:thrust-cpar} we extend our discussion to the case of thrust and the $C$-parameter, and we finally study in Sect.~\ref{subsec:Vb} an observable that smoothly interpolates between thrust and $y_{23}$,
  evaluating the corresponding power corrections.
In Sect.~\ref{sec:summa} we summarise our results. Analytical results for the NLO coefficients for $\tau_2$, $y_{23}$ and $\ktFSR$ are provided in Appendix \ref{app:ABC}, while the exact expression of the three-jet rate with the $k_T$ jet clustering algorithm is reported in Appendix \ref{app:analytic_expr_y23}.

\section{Setup and preliminary investigations}
\label{sec:notation}

We consider the inclusive production of hadrons in $e^+e^-$ annihilation. The LO reaction at parton level is
\begin{equation}
e^+(p_a)+e^-(p_b)\to \gamma^*(q)\to q(p_1)+q(p_2)\, ,  
\end{equation}
where we limit ourselves to consider virtual photon exchange. At NLO the real emission reaction is
\begin{equation}
  \label{eq:proc}
e^+(p_a)+e^-(p_b)\to \gamma^*(q)\to q(p_1)+q(p_2)+g(p_3)\, .  
\end{equation}
The NLO cross section can be written as
\begin{equation}
  \label{eq:snlo1}
   \sigma_{\rm NLO}=\int d\sigma^B+\int d\sigma^R+\int d\sigma^V
\end{equation}
where $d\sigma^B$, $d\sigma^R$ and $d\sigma^V$ are the Born, real and virtual contributions, respectively.
At NLO a slicing method based on a resolution variable $r$ (that we assume to be suitably normalised to make it dimensionless) can be built up by rewriting Eq.~(\ref{eq:snlo1}) as
\begin{equation}
  \label{eq:snlo2}
  \sigma_{\rm NLO}=\int d\sigma^R\theta(r-v)+\left(\int d\sigma^R\theta(v-r)+\int d\sigma^V+\int d\sigma^B\right)\, .
\end{equation}
In Eq.~(\ref{eq:snlo2}) we have split the real contribution into a contribution above and a contribution below a small cut $v$,
using a generic resolution variable $r$.
The first term in Eq.~(\ref{eq:snlo2}) is finite and can be evaluated in $d=4$ dimensions, while
the second term can be evaluated in the small $v$ limit through suitable approximations of the phase space and of the real matrix element in the IR limits. More precisely, one can start from the evaluation of the collinear contributions, and then proceed to add the soft contribution, after subtraction of the soft-collinear terms (see e.g. Ref.~\cite{Buonocore:2023rdw}). Eventually the IR poles from the real contribution below the cut cancel out with those in the virtual contribution and we can write
\begin{align}
  \label{eq:belowcut}
  \int d\sigma^R\theta(v-r)&+\int d\sigma^V+\int d\sigma^B=\nonumber\\
  &=\int d\sigma^B\left(1+\frac{\as(\mu_R)}{\pi}\left(A_r \ln^2v+B_r \ln v+C_r+{\cal O}(v^p)\right)\right)\, .
\end{align}
The explicit form of the coefficients $A_r$, $B_r$ and $C_r$ depends on the choice of the resolution variable $r$, and, in general, also on the Born kinematics.
The power suppressed terms can be neglected if $v$ is sufficiently small. Their structure depends on the observable and we anticipate that they
can be logarithmically enhanced.

In the following we will focus on two resolution variables, the 2-jettiness variable $\tau_2$ \cite{Stewart:2010tn} and the $y_{23}$ resolution variable with the $k_T$ algorithm \cite{Catani:1991hj}. For an event with $n$ final-state partons with momenta $p_1$, $p_2$...$p_n$ the definition of $\tau_2$ is
\begin{equation}
\tau_2=\sum_{k=1}^n {\rm min}\Big\{\frac{2p_k\cdot q_1}{Q^2},\frac{2p_k\cdot q_2}{Q^2}\Big\}
\end{equation}
and depends on the choice of the jet axes $q_1$ and $q_2$. In this paper $q_1$ and $q_2$ are defined by using the JADE clustering algorithm\footnote{We have verified that up to NLO the same clustering history and jet axes are obtained adopting the $k_T$ algorithm with the distances defined as in Eq.~\eqref{eq:ktalg-dij}.}~\cite{JADE:1986kta,JADE:1988xlj}.
Alternative definitions \cite{Moult:2016fqy} directly identify $\tau_2$ with the thrust variable \cite{Farhi:1977sg}, that we will consider in Sec.~\ref{subsec:thrust-cpar}.
The variable $y_{23}$ is instead defined as follows. 
We introduce the distance measure $d_{ij}$ for the $k_T$ algorithm as
\begin{equation}\label{eq:ktalg-dij}
d_{ij}=\frac{2\min\{E_i^2,E_j^2\}(1-\cos\theta_{ij})}{Q^2}\, ,
\end{equation}
where $E_i$ and $\theta_{ij}$ are energies and angular separations defined in the $e^+e^-$ centre-of-mass frame.
The pair with the smallest $d_{ij}$ is clustered and replaced with a pseudo-particle with momentum $p_i+p_j$
and the procedure is repeated until all remaining $d_{ij}$ are larger than some value $y_{\rm cut}$.
The variable $y_{23}$ is defined as the maximum value of $y_{\rm cut}$ for which the event has three jets.
In the NLO case in which only three partons are present, we simply have
\begin{equation}
  y_{23}={\rm min}\{d_{12},d_{13},d_{23}\}\, .
\end{equation}
More generally, we are interested in  observables $r(\{ p_i\}, k)$ whose dependence on the momentum of a single soft emission of momentum $k$, collinear to one of the hard legs of the Born events, can be parametrised as
\begin{equation}\label{eq:obsb}
	r(\{ p_i\}, k) = \left(\frac{k_t^{(\ell)}}{Q}\right)^a e^{-b_\ell \eta^{(\ell)}},
\end{equation}
where $\{ p_i\}$ are the Born momenta and $k_t^{(\ell)}$ and
$ \eta^{(\ell)}(\ge 0)$ denote the transverse momentum and rapidity of
$k$ with respect to the leg $\ell$.  It is easy to show that $\tau_2$
corresponds to the case $a=1,b=1$, while $y_{23}$ corresponds to
$a=2,b=0$.  In order to have an homogeneous scaling in
$k_{t}^{(\ell)}$, in the following we will use
$\yttwth\equiv\sqrt{y_{23}}$.  By limiting ourselves to NLO we can
also consider the variable
\begin{equation}
  \ktFSR=\sqrt\frac{2(p_1\cdot p_3)(p_2\cdot p_3)}{p_1\cdot p_2}
\end{equation}
which represents the transverse momentum of the parton with momentum $p_3$ in the frame in which $p_1$ and $p_2$ are back to back.
\begin{figure}
  \centering
  \includegraphics[scale=0.7]{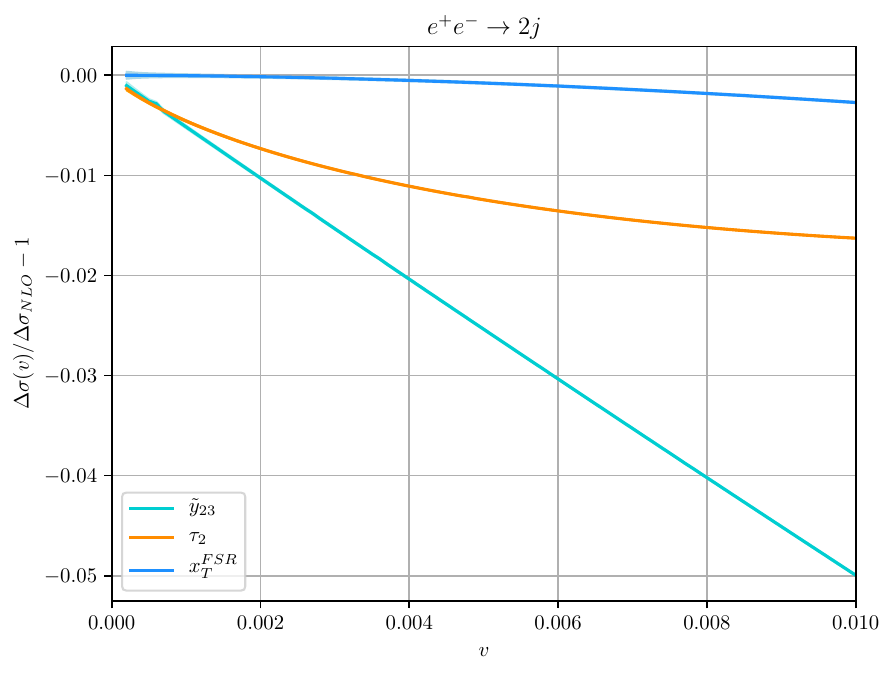}\\
  \caption{Comparison of power suppressed contributions for $\tau_2$, $\yttwth$ and $\xT$}
  \label{fig:slicing}
\end{figure}
We have evaluated the NLO coefficients $A_r$, $B_r$ and $C_r$ in Eq.~(\ref{eq:belowcut}) necessary to carry out the NLO calculation of arbitrary 2-jet observables by using Eq.~(\ref{eq:snlo2}) for the resolution variables $\tau_2$, $\yttwth$ and $\xT\equiv k_T^{\rm FSR}/Q$. The corresponding results are reported in Appendix~\ref{app:ABC}. We can test the quality of the slicing procedure, or, equivalently, the size of power corrections, by plotting the relative deviation of the NLO correction $\Delta\sigma_{\rm NLO}$ from its exact result (see e.g. Ref.~\cite{Ellis:1996mzs}) as a function of $v$.
This is shown in Fig.~\ref{fig:slicing}.

We see that the smallest power corrections are those of the $\xT$ variable, for which the $v$ behavior is consistent with a quadratic dependence.
This is somewhat expected, since this variable strongly resembles the transverse momentum of a colourless system in hadronic collisions\footnote{An $e^+e^-$ observable with a similar behavior \cite{Moult:2019vou} is Energy-Energy correlation (EEC) \cite{Basham:1978bw}.}.
The power corrections for the variable $\tau_2$ are consistent with a logarithmically-enhanced linear behavior.
This could have been expected from the known behaviour of the thrust observable~\cite{DeRujula:1978vmq}, which is equivalent to $\tau_2$ to leading power\footnote{We note, however, that the subleading power corrections for $\tau_2$ and thrust are different, see Sect.~\ref{subsec:thrust-cpar}.}.
On the contrary the $\yttwth$ variable, which represents an effective transverse momentum in the final-state splitting, features purely linear power corrections.
These results are consistent with what observed in Ref.~\cite{Buonocore:2022mle} in the more complicated case of hadronic collisions.
In the following we will check these results through explicit analytic computations, and we will investigate the origin of the different behavior of power corrections.

\section{The calculation}
\label{sec:calculation}

We now focus on the real emission contribution $d\sigma^R$. The three-parton phase space is spanned by five independent variables that can be chosen as
three Euler angles and two of the three energy fractions
\begin{align}
	x_{i} &= \frac{ 2 p_{i} \cdot Q}{Q^2},~~~~~~~~Q=p_a+p_b
\end{align}
that fulfill the energy conservation constraint $x_1 + x_2 + x_3  = 2$.
The variables that we are going to consider are independent of the angles, and, therefore, we can focus on the variables
$x_1$ and $x_2$, whose physical region correspond to the triangle delimited by the lines $x_2=1-x_1$, $x_1=1$ and $x_2=1$ in the $(x_1,x_2)$ plane.
In terms of these variables the resolved real contribution to the cross section, first term of Eq.~\eqref{eq:snlo2}, can be written as
\begin{equation}
  \label{eq:sreal}
        {\sigma}_r^R(v)=\int d\sigma^R\theta(r-v)\equiv\sigma_0 \frac{\as}{2\pi}C_F\, R_r(v)\, ,
\end{equation}
where
\begin{equation}
  \label{eq:real}
R_r(v)=\int_0^1 dx_1 \int_{1-x_1}^1 dx_2f(x_1,x_2)\theta(r(x_1,x_2)-v)\, .
\end{equation}
In Eq.~(\ref{eq:sreal}) $C_F=(N^2_c-1)/(2N_c)$ (with $N_c$ the number of colours), $\sigma_0$ is the LO cross section
\begin{equation}
\sigma_0=\frac{4\pi\alpha^2 N_c\sum_q e_q^2}{Q^2}\, ,
\end{equation}
where the sum is over the quarks $q$ with charge $e_q$ and $\alpha$ is the QED coupling.
The function
\begin{equation}
\label{eq:f}
    f(x_1,x_2) = \frac{x_1^2+x_2^2}{(1-x_1)(1-x_2)}
\end{equation}
in Eq.~(\ref{eq:real}) represents, up to an overall normalisation, the matrix element squared for the process in Eq.~(\ref{eq:proc}).
We recall that the collinear limit $p_3 \parallel p_1$ corresponds to $x_2=1$, while the collinear limit $p_3\parallel p_2$ corresponds to $x_1=1$.
The soft limit $x_{3}\to 0$ is reached in the corner $x_{1,2} \rightarrow 1$.

\begin{figure}
  \centering
  \includegraphics[scale=0.36]{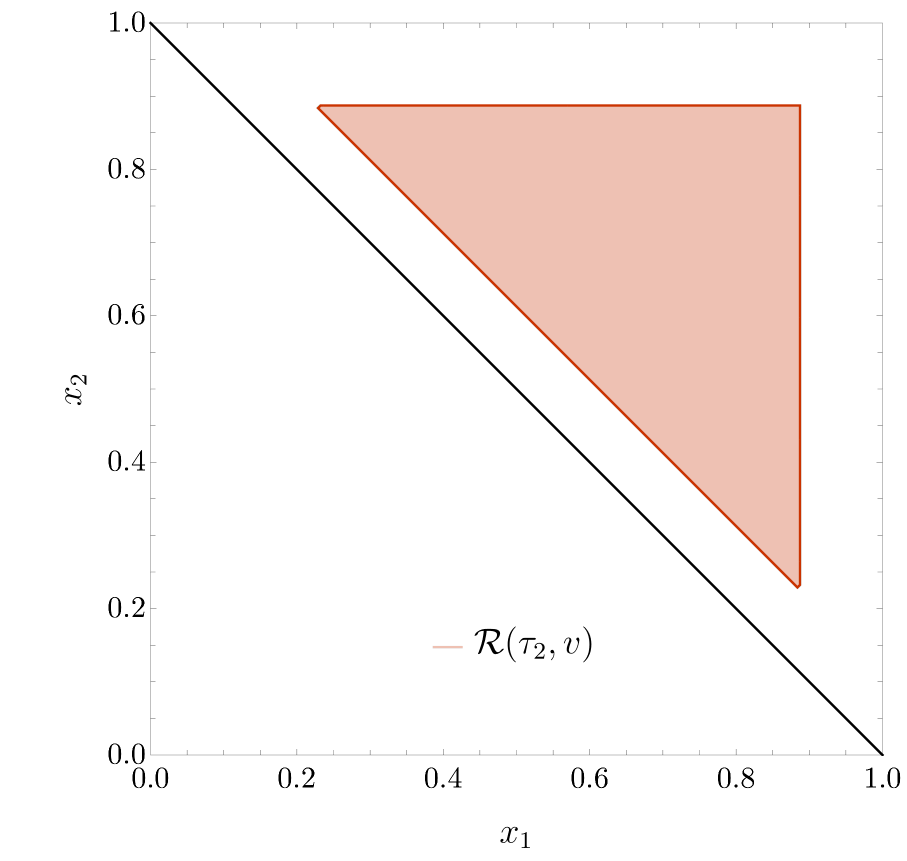}  \includegraphics[scale=0.36]{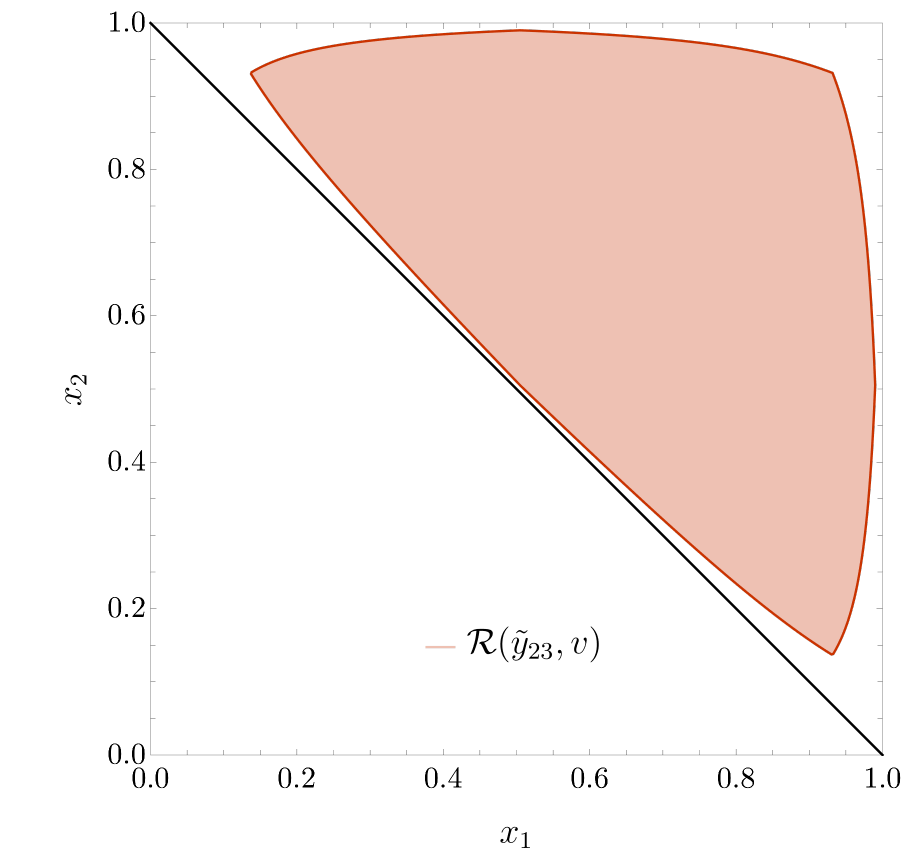} \includegraphics[scale=0.36]{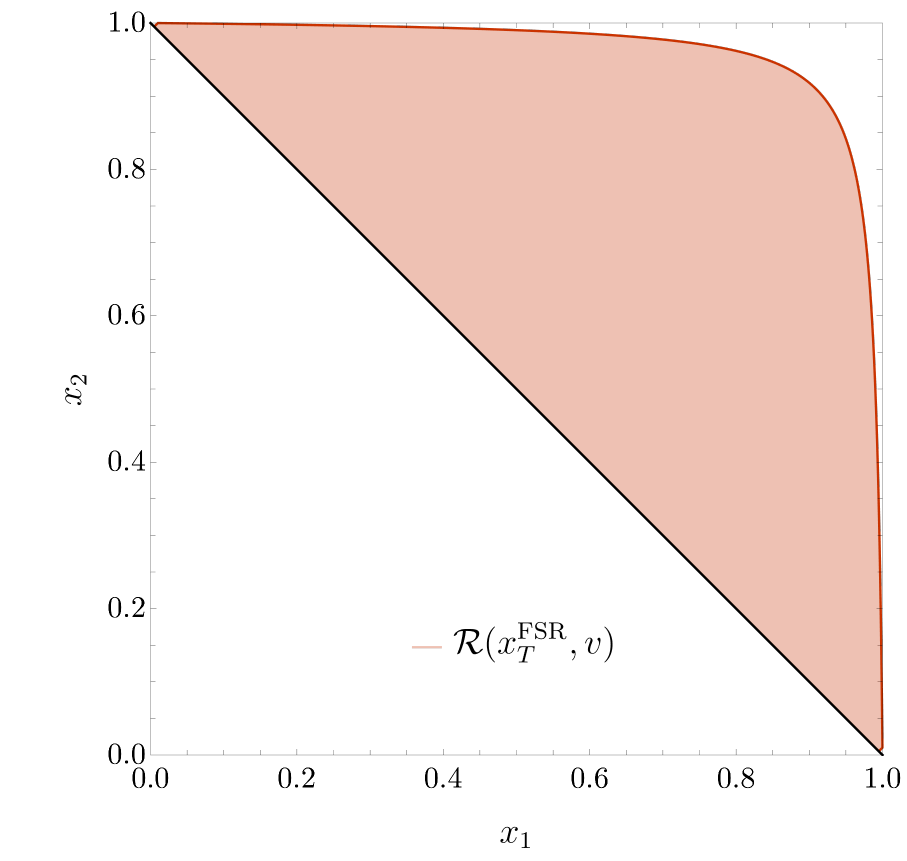}\hfil
  \caption{Regions in the $x_1-x_2$ plane corresponding to the condition $r>v=1/10$ for the variables $\tau_{2}$ (left),  $\yttwth$ (central) and  $x_T^{\rm FSR}$(right).}
  \label{fig:RegsVariables}
\end{figure}

In the case of three partons relevant at NLO, assuming $s_{ij}<s_{ik},s_{jk}$ the jettiness $\tau_2$ variable can be simply written as
\begin{equation}
  \tau_2=x_k(1-x_k)
\end{equation}
where $(i,j,k)$ is an arbitrary permutation of $(1,2,3)$.
We also have
\begin{equation}
  d_{ij}=\frac{\min\{x_i^2,x_j^2\}}{x_i x_j} (1-x_k)
\end{equation}
and
\begin{equation}
  \xT=\sqrt{\frac{(1-x_1)(1-x_2)}{x_1+x_2-1}}\, .
  \end{equation}
It is interesting to study the regions in the $(x_1,x_2)$ plane encompassed by the conditions $r>v$ for the three variables, which are shown in Fig.~\ref{fig:RegsVariables}.

We see that the region $\tau_2>v$ is a triangle, which cuts away the singular regions $x_1\sim 1$ and $x_2\sim 1$ but also a stripe along the line $x_2=1-x_1$.
For the same value of $v$, the region $\yttwth>v$ is larger, and in particular gets closer both to the $x_{1,2}=1$ singular limits as to the non singular region around $x_2=1-x_1$.
The best coverage of the phase space is obtained with the variable $x_T^{\rm FSR}$, which, in particular, fully covers the non singular region around $x_2=1-x_1$.
We can therefore interpret the results in Sec.~\ref{sec:notation} as follows. When the variable $\xT$ becomes small, we are really close to the singular limits
of the matrix element, and the condition $\xT>v$ really cuts only the truly singular region of the $(x_1,x_2)$ plane.
We note that instead, for each values of $v$, a cut on the variable $\yttwth$ leaves out part of a non-singular region along the line $x_2=1-x_1$, which is one of the sources of the different scaling of the power corrections for $\yttwth$.
A cut on the variable $\tau_2$ removes instead a linear stripe along the lines $x_2=1-x_1$, $x_1=1$, $x_2=1$. This can be related to the different dependence on the rapidity of the emission, and, in particular, on the fact that $\tau_2\sim k_T/Q\,e^{-\eta}$. Therefore, a cut $\tau_{2}>v$ induces not only a minimum on the transverse momentum of the radiated parton but also a maximum on its rapidity.
We will see below that this pictorial analysis, which provides us with a qualitative understanding of the scaling of the power corrections, will be confirmed by our explicit calculation.

\subsection{The variable $\xT$}
\label{subsec:xt}

For the variable $\xT$ the real contribution $R_{\xT}(v)$ can be computed exactly in a straightforward way and reads
\begin{equation}
R_{\xT}(v)=\frac{7}{2}+v^2+(3+4v^2+v^4)\ln \frac{v^2}{1+v^2}-2{\rm Li}_2\left(-\frac{1}{v^2}\right)\, .
\end{equation}
In the small-$v$ limit we obtain
\begin{equation}
R_{\xT}(v)= 4\ln^2 v+6 \ln v+\frac{7}{2}+\frac{\pi^2}{3}+4\left(2\ln v-1\right)v^2+{\cal O}(v^4)\, .  
\end{equation}
In this limit the function develops the customary double and single logarithmic contributions.
We also see that, as expected, the power suppressed contributions are quadratic for this variable, consistently to what we have seen in Fig.~\ref{fig:slicing}.

\subsection{The variable $\tau_2$}
\label{subsec:tau2}

We now move to the variable $\tau_2$. From now on, in order to simplify the calculations, we will exploit the
symmetry under the exchange of the quark and antiquark momenta (corresponding to $x_1\leftrightarrow x_2$) and consider
only the integral in the region $x_2>x_1$.

\begin{figure}
  \centering
  \includegraphics[scale=0.4]{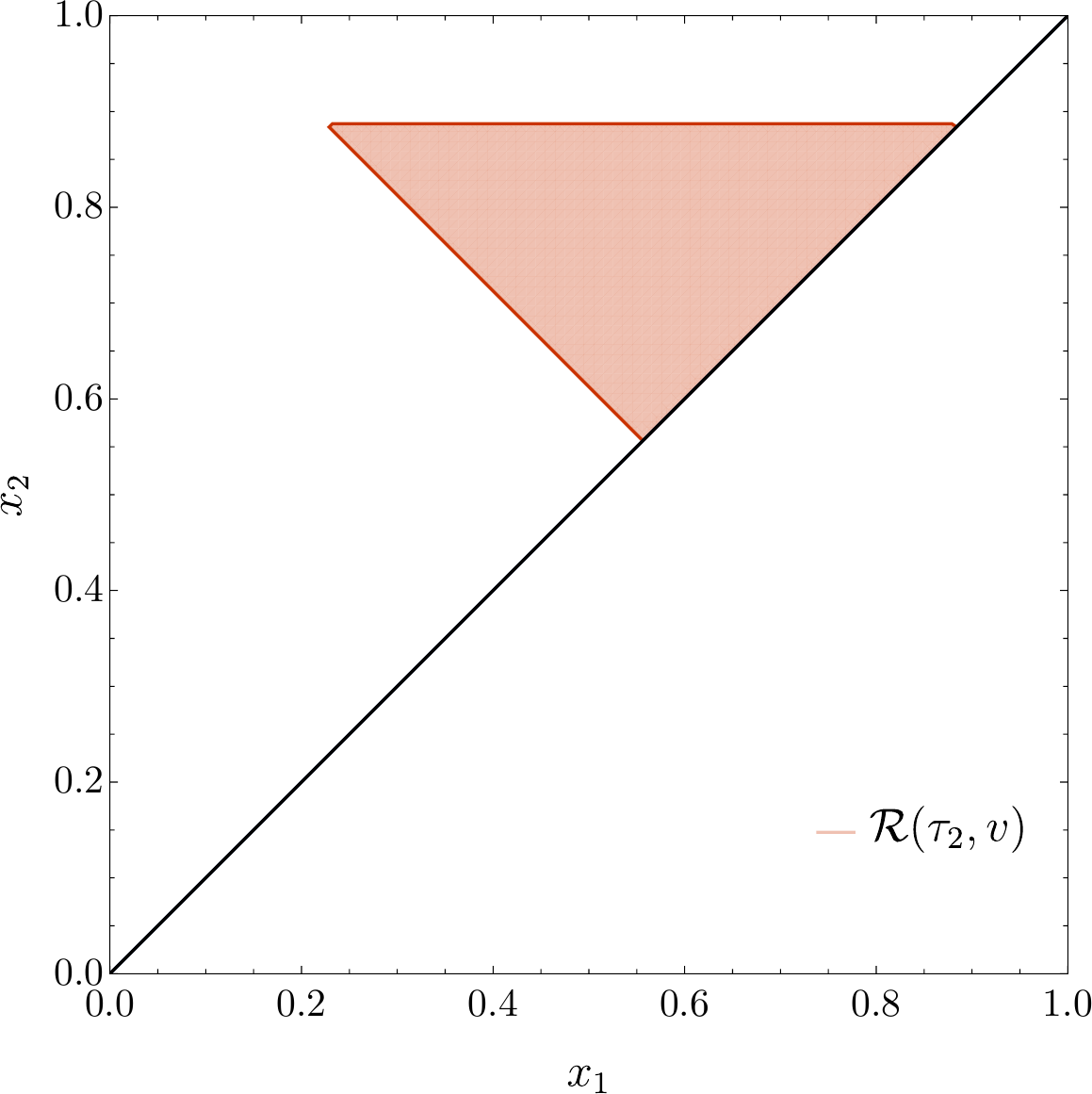}\hspace{1cm}  \includegraphics[scale=0.4]{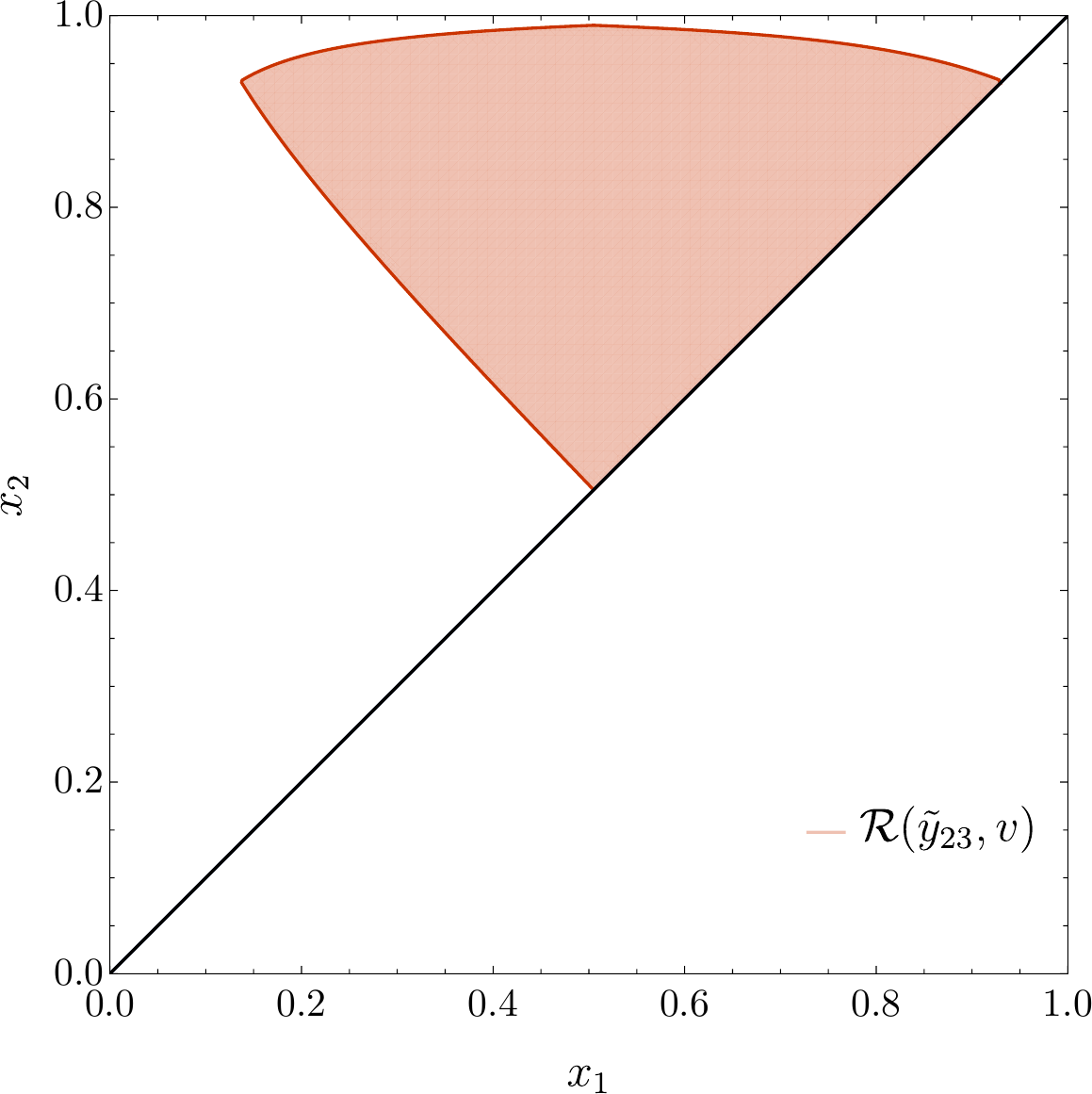}\hfil
  \caption{As in Fig.~\ref{fig:RegsVariables} with the additional constraint $x_{2}>x_{1}$ for $\tau_{2}$ (left) and $\yttwth$ (right). }
  \label{fig:Regs}
\end{figure}

In the $(x_1,x_2)$ plane, the cut $\tau_2>v$ defines a triangular region as shown in the left panel of Fig.~\ref{fig:Regs}. A similar contour is obtained for the case of the thrust event shape \cite{Farhi:1977sg}, and also for the three-jet region defined by the JADE clustering algorithm~\cite{JADE:1986kta}.
We can extend our calculation for this class of observables by considering the following parametrisation of the region
 \begin{align}
   {\cal R}(v) = \bigg\{2 u < x_1< \frac{1}{2}(1+u),  1 - x_1 + u < x_2 < 1-u
      \vee \frac{1}{2}(1+u) < x_1, x_1 < x_2 < 1 - u
   \bigg\},\;
\end{align}
where, for example, $u(v) = v$ for thrust and $u(v) = \frac{1}{2}(1-\sqrt{1-4v}) = v + \mathcal{O}(v^{2})$ for $\tau_{2}$.
The real contribution is obtained by integrating the function $f(x_1,x_2)$ in the above region, namely
\begin{align}
  \label{eq:pc-triangle-master}
R_r(v)&=2\int _{{\cal R}(v)} \!\!\mbox{d}x_1\mbox{d}x_2 \, f(x_1,x_2)  \nonumber\\
&=\frac{5}{2} - \frac{\pi ^2}{3} + 2\ln^2\left(\frac{1-u}{u}\right) + \left(6 u-3\right) \ln \left(\frac{1-2u}{u}\right) - 6 u- \frac{9 u^2}{2} + 4\text{Li}_2\left(\frac{u}{1-u}\right)\,.
\end{align}
We focus here on the case of 2-jettiness variable and postpone the discussion on thrust to Sec~\ref{subsec:thrust-cpar}.
For $r=\tau_{2}$ we have $u(v) = \frac{1}{2}(1-\sqrt{1-4v})$ and
\begin{align}
  \label{eq:tau2exact}
 R_{\tau_{2}}(v) &=  -\frac{11}{4}-\frac{\pi ^2}{3}+2 \ln ^2\left(\frac{2}{1-\sqrt{1-4v}}-1\right) -3 \sqrt{1-4 v} \ln \left(\frac{2}{1-\sqrt{1-4v}}-2\right) \notag \\ &+ \frac{9v}{2}+\frac{21}{4} \sqrt{1-4 v} + 4 \text{Li}_2\left(-\frac{2 v+\sqrt{1-4 v}-1}{2 v}\right) \notag \\
  &=  2 \ln^{2}{v} + 3 \ln{v} +\frac{5}{2} -\frac{\pi^{2}}{3} +
  v(7+2\ln{v}) + v^{2}\left(5+6\ln{v} \right) + \mathcal{O}(v^{3})\,.
\end{align}
We notice that the subleading power correction is linear and is logarithmically-enhanced, consistently to what we have observed in Fig.~\ref{fig:slicing}.

\subsection{The variable $\yttwth$}
\label{subsec:y23}

A similar analysis can be carried for $\yttwth$.
The region $\yttwth > v$ is given by (see right panel of Fig.~\ref{fig:Regs})
\begin{eqnarray}\label{eq:S2y23}
	{\cal R}(\yttwth; v) \!\!\! &=& \!\!\! \bigg\{ \frac{v}{2}\sqrt{8+v^{2}} - \frac{v^{2}}{2} < x_1< \frac{1}{2} + \frac{v^{2}}{2}, 1-x_{1} + v^{2}\frac{1-x_{1}}{x_{1}-v^{2}}  < x_2 < 1 - v^{2}\frac{1-x_{1}}{x_{1}-v^{2}} \notag\\
&&\hspace{-2.3cm} \vee \; \frac{1}{2} + \frac{v^{2}}{2} < x_1< 1 -\frac{v}{4}\sqrt{8+v^{2}} + \frac{v^{2}}{4} , x_{1} < x_2 < \frac{3}{2}-\frac{x_{1}}{2} - \frac{1}{2}\sqrt{1+x_{1}(x_{1}-2+4v^{2})}\bigg\}\;.
\end{eqnarray}
The integral can be computed analytically, but the final result (which corresponds to the LO 3-jet rate with the $k_T$ algorithm) is less compact
than that for $\tau_2$ and is reported in App.~\ref{app:analytic_expr_y23}. We find agreement with the result of Ref.~\cite{Brown:1991hx}, provided a typo therein is corrected.
By expanding in $v$, we observe
that the power correction is again linear in $v$, but does not contain
any logarithmic enhancement:
\begin{equation}
  R_{\yttwth}(v) = 4\ln^{2}{v} + 6 \ln{v} +\frac{5}{2}-\frac{\pi^{2}}{6} + 6\ln{2} + \left(\!4\ln\left(1 \!+ \! \sqrt{2}\right)  \!- 8\sqrt{2}\right) v+\left(5\! - 18\! \ln 2 - \! 8 \ln v\right)v^2+{\cal O}(v^3).
  \label{eq:risy23t}
\end{equation}
The first occurrence of a logarithmically-enhanced term appears at ${\cal O}(v^2)$.

\subsection{Comparison between $\tau_{2}$ and $\yttwth$}
\label{subsec:comparison-tau2-y23}

\begin{figure}
  \centering
  \includegraphics[scale=0.4]{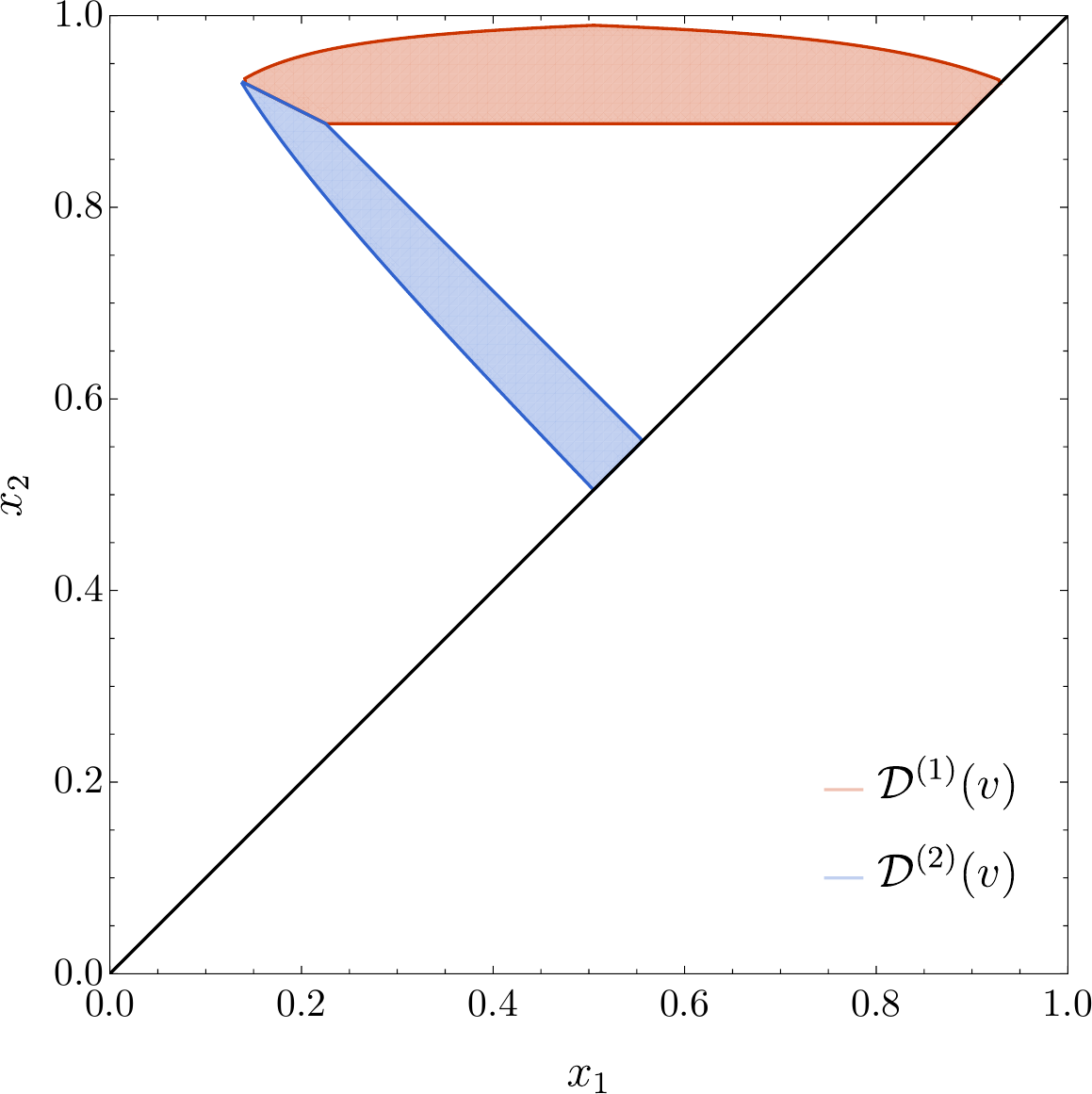}\hspace{1cm}  \includegraphics[scale=0.4]{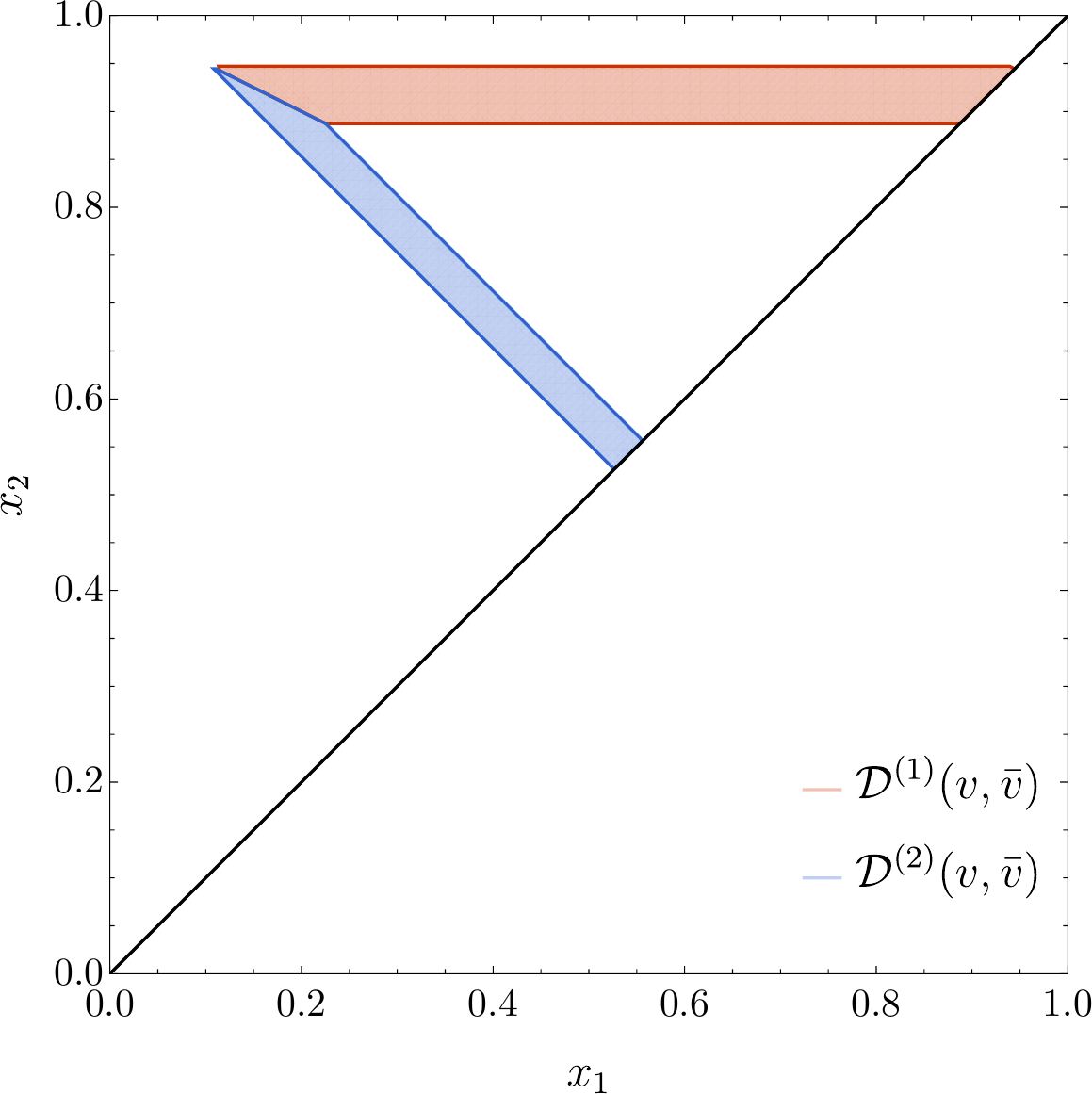}\hfil
  \caption{Regions ${\cal D}^{(1,2)}(v)$ (left) and ${\cal D}^{(1,2)}(v,\bar{v})$ (right) in the $x_1-x_2$ plane.}
  \label{fig:DRegs}
\end{figure}

Although we could perform the two calculations analytically, thereby obtaining
the full tower of power corrections at order $\as$, this analysis does not
shed light on the physical origin of the power corrections nor on the observed
difference between the two cases. To gain further insight, we compare the
regions ${\cal R}(\tau_{2};v)$ and ${\cal R}(\yttwth;v)$ associated with the two variables for
the same value of the parameter $v$. The situation is illustrated in
Fig.~\ref{fig:DRegs}. We observe that the region ${\cal R}(\tau_{2};v)$ is included in
${\cal R}(\yttwth;v)$ and we focus on the region ${\cal D} = {\cal R}(\yttwth;v)\backslash {\cal R}(\tau_{2};v)$. Since the variable $\yttwth$ does not feature logarithmically-enhanced power corrections, the integral of the matrix element in the region ${\cal D}$ must give rise to the same logarithmically-enhanced power corrections of $\tau_{2}$, but with an opposite sign.
In order to identify the phase space regions responsible for the presence of logarithmically-enhanced power corrections, we further split the region ${\cal D}$
into two subregions ${\cal D}^{(1)}$ and ${\cal D}^{(2)}$ by connecting the two corners by a
straight line, whose equation is simply given by $1-x_{1}/2-x_{2}=0$, as shown in
Fig.~\ref{fig:DRegs}.
We perform the integration over the two regions. The results expanded up to $\mathcal{O}(v)$ read
\begin{align}
  2 \int_{{\cal D}^{(1)}}d x_{1} d x_{2} f(x_{1},x_{2}) &= 2\ln ^2{v} + 3 \ln{v}  + \frac{\pi ^2}{6} + 6\ln 2 \notag \\&+ v \left(-7-8
  \sqrt{2}+8\ln\left(1+\sqrt{2}\right)\right) + \mathcal{O}(v^{2})  \; ,
\end{align}
and
\begin{equation}
    2 \int_{{\cal D}^{(2)}}d x_{1} d x_{2} f(x_{1},x_{2}) = -4 v\ln\left(1+\sqrt{2}\right)  - 2 v \ln v + \mathcal{O}(v^{2}) \;,
\end{equation}
respectively.

The main result of the above analysis is that the logarithmically
enhanced linear power correction comes entirely from the region ${\cal D}^{(2)}$,
where, as expected, it appears with opposite sign with respect to the one
present for the $\tau_{2}$ variable\footnote{We note that the way we separate the regions ${\cal D}^{(1)}$ and ${\cal D}^{(2)}$, which naturally follows from the definitions of $\yttwth$ and $\tau_2$, is crucial to our observation. A different splitting in the decomposition may shift the logarithmically-enhanced contribution between the two.}. This region corresponds to physical
configurations in which the gluon is hard and recoils against a collinear and/or
soft quark-antiquark pair. In fact, we are far from configurations in phase
space where the real matrix element develops IR singularities, and so
the contribution stemming from the region ${\cal D}^{(2)}$ is a pure power correction.
The fact that the logarithmically-enhanced power corrections are entirely due to the non singular region close to the $x_2 = 1- x_1$ line is non trivial. As we will see in the following when considering the case of thrust and $C$-parameter, the absence of logarithmically-enhanced power corrections in the ${\cal D}^{(1)}$ region is a peculiar characteristic of $\tau_{2}$.

Having identified the phase space region responsible for the logarithmically-enhanced power corrections, we would like to confirm that their origin is purely kinematical. To this end, we turn our attention to the matrix
element and we consider its approximation in the singular limits. The only
singular limit approached in the region ${\cal D}^{(2)}$ is the collinear limit $x_{2}\to 1$,
where the momentum of the gluon becomes parallel to the one of the quark. We
perform, then, the integration over ${\cal D}^{(2)}$ of the matrix element in this limit, which implies replacing the function $f(x_1,x_2)$ in Eq.~(\ref{eq:f}) with
the leading term $f_{\rm coll}^{(0)}(x_1,x_2)$ of the collinear expansion of the matrix element
\begin{equation}\label{eq:collinearlimit}
  f (x_{1},x_{2}) = \frac{1+x_{1}^{2}}{(1-x_{1})(1-x_{2})} - \frac{2}{1-x_1}+ \mathcal{O}(1-x_2) \equiv f_{\rm coll}^{(0)} (x_{1},x_{2}) + f_{\rm coll}^{(1)} (x_{1},x_{2}) + \mathcal{O}(1-x_2)\, .
\end{equation}
The result reads
\begin{equation}
    2 \int_{{\cal D}^{(2)}}d x_{1} d x_{2} f^{(0)}_{\rm coll}(x_{1},x_{2}) = v \left( 1 + 2\ln{2} - 4 \ln\left(1+\sqrt{2}\right)    -2 \ln{v} \right) + \mathcal{O}(v^{2})  \;.
\end{equation}
We observe that the collinear approximation of the matrix element is
sufficient to correctly recover the logarithmically-enhanced linear
power correction.  Furthermore, we checked that this remains true also
by setting $x_1 = 0$ in the expression of the collinear matrix element
$f^{(0)}_{\rm coll}$, i.e. by considering the limit in which the quark
becomes soft.

The picture that emerges is that this contribution is a consequence of removing
a phase space region which is
  non-singular but extends itself into the collinear limits, because of
  the cut on $\tau_{2}$. By contrast, we notice that integrating down to the
$\yttwth$ contour does not lead to the appearance of a similar
logarithmically-enhanced linear power correction. We associate this result to
the fact that the phase space volume removed by imposing the cut on $\yttwth$
scales quadratically with $v$ whereas it scales linearly for the case of
$\tau_{2}$. In turn, the different profile of the contour is a consequence of
the different rapidity dependence of the
variable in the collinear limit, i.e. the
exponent $b$ in Eq.~\eqref{eq:obsb}. Before moving forward, we
complete the above discussion repeating the same exercise replacing the
${\cal R}(\yttwth,v)$ region with another $\tau_{2}$ region ${\cal R}(\tau_{2},\bar{v})$ with
  $\bar{v} < v$, as shown in the right panel of Fig.~\ref{fig:DRegs}. Performing the integration over the region ${\cal D}^{(2)}(v,\bar{v})$, we obtain
 \begin{equation}
  2 \int_{{\cal D}^{(2)}(v,\bar{v})}d x_{1} d x_{2} f(x_{1},x_{2}) = 2\bar{v} \ln {\bar{v}} - 2 v \ln v + \mathcal{O}(v^{2},\bar{v}^{2})\;,
\end{equation}
 which is consistent with our expectation that this region is the one responsible for the logarithmically-enhanced linear power correction. We note that the integral in the region ${\cal D}^{(2)} (v, \bar v)$ does not give rise to linear, non-logarithmically-enhanced power corrections, which are thus entirely contained in the region ${\cal D}^{(1)} (v, \bar v)$.

In conclusion, we have shown that, for the case of $\tau_2$, the logarithmically-enhanced power correction is a pure phase space effect. The simplicity of this result is observable dependent, as we will discuss in the following section. In fact, one generally expects contributions to the power correction also stemming from the expansion of the real matrix element beyond the leading power. Nonetheless, we anticipate here that in the non-singular region close to the boundary $x_{2} = 1-x_1$ the collinear approximation of the matrix element is sufficient to capture the logarithmically-enhanced power correction also for the variables considered in the next section.

\subsection{Thrust and $C$-parameter}
\label{subsec:thrust-cpar}
In this section we will study the cases of thrust \cite{Farhi:1977sg} and of C-parameter~\cite{Parisi:1978eg,Fox:1978vu,Donoghue:1979vi}.
We start with thrust $T$ and consider the observable $1-T$ .
By using the energy fractions we can write
\begin{equation}
 1- T= \min\{1-x_{1},1-x_{2},1-x_{3}\}\, .
\end{equation}
The exact result for the cumulative cross section is given by Eq.~\eqref{eq:pc-triangle-master} with $u(v)=v$, which reproduces the known result in the literature~\cite{DeRujula:1978vmq}.
Expanding in $v$ we obtain
\begin{equation}
  \label{eq:RT}
  R_{1-T}(v) = 2 \ln^{2}{v} + 3 \ln{v} +\frac{5}{2} -\frac{\pi^{2}}{3} +
  2v(2-\ln(v)) - v^{2}\left(\frac{7}{2}-2\ln{v} \right) + \mathcal{O}(v^{3})\,.
\end{equation}
Comparing Eq.~(\ref{eq:RT}) with Eq.~(\ref{eq:tau2exact}) we see that
the expansion of $R_{\tau_2}(v)$ and $R_{1-T}(v)$ coincides at the leading power ${\cal O}(v^0)$, including the constant term. This is not unexpected, since these variables behave exactly in the same way in the relevant IR limits. However, the subleading power corrections are different, as the two variables start to depart from each other going beyond the soft and collinear approximations. In particular, the subleading power corrections are logarithmically enhanced in both cases but with a different coefficient. We have repeated the analysis of Sect.~\ref{subsec:comparison-tau2-y23} for thrust, studying the contribution to subleading power corrections from the regions ${\cal D}^{(1)}$ and ${\cal D}^{(2)}$. We find that, contrary to what happens for $\tau_2$, the subleading-power logarithmic term does not originate only from ${\cal D}^{(2)}$ but there is also a contribution from ${\cal D}^{(1)}$. As anticipated, the contribution from ${\cal D}^{(2)}$ can be obtained through a collinear approximation of the matrix element, extended into the non-singular region, and is identical to that of $\tau_2$. The contribution of ${\cal D}^{(1)}$ can be exactly obtained from a collinear approximation of the matrix element including both the leading and the next-to-leading power contributions\footnote{We have also checked that a subleading soft approximation (see e.g. Ref~\cite{Agarwal:2023fdk}) is able to capture the correct coefficient of the logarithmically-enhanced power correction in the region ${\cal D}^{(1)}$.} $f_{\rm coll}^{(0)}$ and $f_{\rm coll}^{(1)}$ in Eq.~\eqref{eq:collinearlimit}, and, combined with the ${\cal D}^{(2)}$ contribution, leads to the result reported in Eq.~(\ref{eq:RT}). Our result and the associated interpretation of the origin of the logarithmically-enhanced subleading power correction for thrust is in perfect correspondence with the analysis performed in Ref.~\cite{Moult:2016fqy} in the SCET framework.

We now move to the case of the $C$-parameter.
For final-state massless particles the $C$-parameter can be defined as
\begin{equation}
C=3-\frac{3}{2}\sum_{i,j}\frac{(p_i\cdot p_j)^2}{(p_i\cdot q)(p_j\cdot q)}\, .
\end{equation}
The two-jet limit corresponds to $C\to 0$ and in this limit the $C$ parameter and thrust are related by
\begin{equation}
  C=6(1-T)\, .
\end{equation}
This relation holds up to next-to-leading logarithmic accuracy \cite{Catani:1998sf}.
In the following we will consider the variable $c=C/6$ which can be written in terms of the energy fractions as
\begin{equation}
  c=\frac{(1-x_1)(1-x_2)(1-x_3)}{x_1 x_2 x_3}\, .
\end{equation}
The evaluation of the cumulative cross section $R_c(v)$ in this case is more complicated and involves elliptic integrals \cite{Gardi:2003iv,Agarwal:2023fdk}. In the $v\to 0$ limit we find
\begin{equation}
\label{eq:resC}
R_c(v)=2\ln^2 v+3\ln v+\frac{5}{2}-\frac{2}{3}\pi^2+v(7-4\ln v)+{\cal O}(v^2)\, .
\end{equation}
We see that the logarithmic terms are the same as those in Eqs.~(\ref{eq:RT}) and (\ref{eq:tau2exact}), but the constant term is different.
We also see that the subleading power correction is logarithmically-enhanced\footnote{We note that our result for the coefficients of the subleading power correction differs from the expansions reported in Eq.~(B.10) of Ref.~\cite{Gardi:2003iv} and in Eq.~(89) of Ref.~\cite{Agarwal:2023fdk}, while it agrees with a numerical evaluation of the full expression in terms of elliptic integrals.}, with a different coefficient
with respect to that of $(1-T)$ and $\tau_2$. By repeating the analysis in the corresponding regions ${\cal D}^{(1)}$ and ${\cal D}^{(2)}$, we observe the same pattern as for thrust.
Summarising we have 
\begin{equation}
  2 \int_{{\cal D}^{(2)}(v)}d x_{1} d x_{2} f(x_{1},x_{2})  \sim 2 \int_{{\cal D}^{(2)}(v)}d x_{1} d x_{2} f^{(0)}_{\rm coll}(x_{1},x_{2})  \sim - 2 v \ln v \;,
\end{equation}
valid for both thrust and $C$-parameter and
 \begin{equation}
   2 \int_{{\cal D}^{(1)}(v)}d x_{1} d x_{2} f(x_{1},x_{2})  \sim \begin{cases}
                                                                    + 4 v \ln v ;, \quad \text{for } 1-T\\
                                                                    + 6 v \ln v ;, \quad \text{for } c\\
                                                                  \end{cases}\;.
\end{equation}
In the above formulae, with the symbol $\sim$ we mean that we are restricting the result to the logarithmically-enhanced subleading power correction. As anticipated, in the region ${\cal D}^{(2)}$ the latter has a common origin and the same coefficient for all three considered variables.

\subsection{The variable $r_b$}
\label{subsec:Vb}

In this section we investigate in more detail the presence of
logarithmically-enhanced power corrections for a variable of the kind
of Eq.~(\ref{eq:obsb}) with a generic $b$
exponent~\cite{Berger:2003iw}. Observables of such kind have
been considered in Ref.~\cite{Banfi:2004yd} and were recently studied
in order to assess the logarithmic accuracy of Monte Carlo parton
showers~\cite{Dasgupta:2020fwr}. The motivation of introducing such
family of shower ordering variables is related to their different
coverage of the Lund plane~\cite{Andersson:1988gp}, which, in
combination with an appropriate treatment of the recoil of the
emission, may ultimately affect the possibility to achieve
next-to-leading logarithmic accuracy or beyond.

Based on the discussion in Sect.~\ref{subsec:comparison-tau2-y23}, and, in particular, on our
observation that $\tau_2$ corresponds to the case $a=1$ and $b=1$ in Eq.~(\ref{eq:obsb}),
the most natural definition of such general observable for our NLO analysis
would be obtained through an appropriate combination of $\yttwth$ and $\tau_2$.
However, we have seen that $\tau_2$ is quite special, since with our definition of the jet axes the logarithmically-enhanced power correction originates only in the region ${\cal D}^{(2)}$. We therefore use $1-T$ instead of $\tau_2$.
We define the class of observables
\begin{equation}\label{eq:gen_obs_def}
    r_b = (1-T)^b \, \yttwth^{1-b},
\end{equation}
that smoothly interpolates between the two limits $b=0$ ($\yttwth$) and $b=1$ ($1-T$).
These observables admit a compact expression as a function of $x_i$, facilitating our analysis in the $(x_1,x_2)$ plane.
We note that these observables are not recursive infrared collinear safe~\cite{Banfi:2003je} being a combination of two recursive infrared collinear safe observables but with a different $b$~\cite{Banfi:2004yd}.
This, however, is not an issue in our case, since we are looking for an observable that is sufficiently simple to allow for the evaluation of the leading power corrections in analytic form.
We have computed the cumulative cross section $R_{r_b}$ for this observable, including subleading power corrections. We find:
\begin{align}
  \label{eq:rb-main-result}
  R_{r_b}(v)&=\frac{2}{1+b}\left(2\ln^2 v+3\ln v\right)+\frac{5}{2}-(1+b)\,\frac{\pi^2}{6}+6\,\frac{1-b}{1+b}\,\ln 2+\left[\,\frac{2^{\frac{5+b}{2}}b}{1+b}\right.\nonumber\\
  &\left.+4B_{1/2}\left(-\frac{1+b}{2},0\right)-2B_{1/2}\left(\frac{1-b}{2},0\right)\right]v
  +\Bigg[4B_{1/2}\left(\frac{b-1}{b+1}, 0\right) - 4B_{1/2}\left(\frac{2 b}{1 + b},0\right)\nn\\
  &+\frac{\Gamma \left(\frac{b-1}{b+1}\right) \left(4
   \left(b^4+3b^3+6b^2+b+1+\frac{b(b^3-7b^2+3b+3)}{b+1}
   B_{\frac{1}{2}}\left(\frac{b-1}{b+1},\frac{2}{b+1}\right)\right)-4 b^{\frac{2+b}{1+b}}
   (b+1)^2\right)}{(b+1)^3\, \Gamma \left(\frac{2 b}{b+1}+1\right)}\nonumber\\
  &+\frac{5 b^2+ 6 b-3}{(1+b)^2}\left(\psi\left(\frac{b}{1+b}\right)-\psi\left(\frac{1+3b}{2(1+b)}\right)\right)\Bigg]v^{\frac{2}{1+b}}
  +{\cal O}\left( v^2\right)\, ,
\end{align}
where the incomplete Beta function is defined as
\begin{equation}
  B_{z}(a,b) = \int_{0}^{z} dt \, t^{a-1}(1-t)^{b-1}\,.
\end{equation}
Eq.~\eqref{eq:rb-main-result} shows that the structure of the power corrections
of the generic observable $r_{b}$ is richer, since it contains an additional tower of
non-rational power corrections of the type $\left(v^{2/(1+b)}\right)^{k}$.
The presence of non-rational power corrections for $b \in (0,1)$ is consistent with the findings of Ref.~\cite{Budhraja:2019mcz} in the context of the all-order resummation of angularities in SCET.

We see that the subleading power correction for $r_b$ does not display explicit logarithmic enhancements, similarly to what happens for $\yttwth$. It is easy to check that in the limit $b\to 0$ the linear term in Eq.~\eqref{eq:rb-main-result} reproduces the linear term for $\yttwth$ in Eq.~(\ref{eq:risy23t}). On the other hand, the rather complex analytical structure of $r_b$ leads to a log-like behaviour for values of $b \lesssim 1$. In the limit $b\to 1$ the coefficient of the linear power correction is divergent, and combined with the $b\to 1$ limit of the ${\cal O}(v^{2/(1+b)})$ term, reproduces the subleading power correction for the $1-T$ variable in Eq.~(\ref{eq:RT}).

\section{Summary}
\label{sec:summa}

In this paper we have considered subleading power corrections to event shape variables in $e^+e^-$ collisions. 
We have started from the jettiness variable $\tau_2$
and the $y_{23}$ resolution variable for the $k_T$ jet clustering algorithm.
We have computed the necessary ingredients to use these variables as slicing variables to evaluate generic $e^+e^-\to 2$ jet observables at NLO.
Both variables are affected by linear power corrections in the two-jet limit. In the case of jettiness
the power correction is logarithmically-enhanced, while for $y_{23}$ this is not the case. We have also
considered a toy variable $\ktFSR$, which can be defined at NLO as the transverse momentum of the gluon with respect to the quark-antiquark pair.
This variable resembles the transverse momentum of a colourless system in hadron collisions and shows quadratic power corrections.

We have analytically computed the
cumulative cross section for these observables, and discussed the origin of the different power corrections.
Our main observation is that these variables cover the phase space in different ways, and that the different power corrections can be attributed to how they cut the singular region in the $(x_1,x_2)$ plane.
We have also shown that, with our definition, the logarithmically-enhanced power correction for $\tau_2$ can be obtained through a collinear approximation of
the matrix element that is extended to the non-singular region. We have then extended our analysis to thrust and to the $C$-parameter, presenting the expression
of the subleading-power correction. In this case, the logarithmic contribution does not stem only from the collinear approximation extended to the non-singular region, but also from a subleading power collinear expansion of the matrix element.

We finally considered a class of variables $r_b$
that depend on a continous parameter giving different weight to central and forward emissions.
Similar variables have been considered in recent studies of the logarithmic accuracy of parton showers \cite{Dasgupta:2020fwr}.
We have defined these variables through a smooth interpolation between $1-T$ and $y_{23}$.
We have shown that these variables have a non-trivial structure of non-rational power corrections, as observed for angularities~\cite{Budhraja:2019mcz}, and we have evaluated the ${\cal O}(v)$ and ${\cal O}\left(v^{(2/(1+b)}\right)$ terms in this expansion. We have shown that no logarithmically-enhanced correction emerges at ${\cal O}(v)$ and at order ${\cal O}(v^{(2/(1+b)})$ for $b < 1$.

Recent studies of subleading power corrections to event shape variables concentrated on the thrust and jettiness variables and were mostly carried out within Soft Collinear Effective Theory \cite{Moult:2016fqy,Boughezal:2016zws,Moult:2018jjd,Moult:2019mog,Moult:2019uhz,Beneke:2020ibj,Beneke:2022obx}. Our results extend these findings to $y_{23}$, to the $C$-parameter and to the new class of variables $r_b$, offering a different perspective on the structure of power corrections and can also be useful to understand and improve the efficiency of non-local subtraction schemes.
The findings of this work suggest a connection between the rapidity dependence of the observable and the scaling of the leading power corrections. Specifically, we found that observables which do not depend on the rapidity of the emission do not feature linear logarithmically-enhanced power corrections at NLO. As a consequence, for such observables, the onset of logarithmically-enhanced linear power corrections, which is expected on general grounds, starts from the next-to-next-to-leading order.

\vskip 0.5cm
\noindent {\bf Acknowledgements}

\noindent We would like to thank Stefano Catani, Pier Monni and
Gherardo Vita for helpful discussions and comments. This work is
supported in part by the Swiss National Science Foundation (SNSF)
under contracts 200020$\_$188464 and PZ00P2$\_$201878 and by the UZH
Forschungskredit Grant FK-23-098. The work of LB is funded by the
European Union (ERC, grant agreement No. 101044599, JANUS). Views and
opinions expressed are however those of the authors only and do not
necessarily reflect those of the European Union or the European
Research Council Executive Agency. Neither the European Union nor the
granting authority can be held responsible for them.

\appendix

\section{NLO coefficients for $\tau_2$, $\yttwth$ and $\xT$}
\label{app:ABC}

In this appendix, we report the explicit expressions of the leading power coefficients $A_{r}$, $B_{r}$ and $C_{r}$ entering Eq~\eqref{eq:belowcut}, which we write again here for ease of the reader
\begin{align}
  \int d\sigma^R\theta(v-r)&+\int d\sigma^V+\int d\sigma^B=\nonumber\\
  &=\int d\sigma^B\left(1+\frac{\as(\mu_R)}{\pi}\left(A_r \ln^2v+B_r \ln v+C_r+{\cal O}(v^p)\right)\right)\, ,
\end{align}
for the three resolution variables considered in the main text. The calculation proceeds along the line of Ref.~\cite{Buonocore:2023rdw}, and, explicitly, it requires the computation of the observable-dependent NLO quark-jet and soft functions, and the observable-independent finite remainder of the one-loop virtual amplitude.

\begin{itemize}
  \item $r = \xT$:
        \begin{align}
          A_{\xT} &= -2C_{F} ,\quad  B_{\xT} = -3C_{F} , \nonumber \\
          C_{\xT} &=  C_{F}\left( \underbrace{\frac{\pi^{2}}{2}-4}_{\text{Virtual}} + \underbrace{3 - \frac{5\pi^{2}}{6}}_{2\times \text{Jet}_{q}} + \underbrace{\frac{\pi^{2}}{6}}_{\text{Soft}} \right) = -C_{F}\left( \frac{\pi^{2}}{6} + 1 \right)
        \end{align}
  \item $r = \tau_{2}$:
        \begin{align}
          A_{\tau_{2}} &= -C_{F} ,\quad  B_{\tau_{2}} = -\frac{3}{2}C_{F} , \nonumber \\
          C_{\tau_{2}} &=  C_{F}\left( \underbrace{\frac{\pi^{2}}{2}-4}_{\text{Virtual}} + \underbrace{\frac{7}{2} - \frac{\pi^{2}}{2}}_{2\times \text{Jet}_{q}}  +\underbrace{ \frac{\pi^{2}}{6}}_{\text{Soft}} \right) = C_{F}\left( \frac{\pi^{2}}{6} - \frac{1}{2}  \right)
        \end{align}
  \item $r = \yttwth$:
        \begin{align}
          A_{\yttwth} &= -2C_{F}, \quad B_{\yttwth}= -3C_{F}, \nonumber \\ C_{\yttwth} &= C_{F}\left( \underbrace{\frac{\pi^{2}}{2}-4}_{\text{Virtual}} + \underbrace{ \frac{7}{2} - \frac{\pi^{2}}{2} - 3\ln2}_{2\times \text{Jet}_{q}} +\underbrace{ \frac{\pi^{2}}{12}}_{\text{Soft}} \right) = C_{F}\left( \frac{\pi^{2}}{12} - \frac{1}{2}  - 3\ln2  \right)
        \end{align}
\end{itemize}

\section{The variable $\yttwth$: exact analytic result}
\label{app:analytic_expr_y23}
In this appendix we report the exact expression for the $\yttwth$ variable\footnote{The formula is also given in the ancillary mathematica notebook {\tt res\_y23.nb} in the arXiv submission.}:
\begin{align}
R_{\yttwth}(v) &=
\frac{64 \ln {v} v^6}{\left(-9 v-3 \sqrt{2} t+u\right)^2}+\frac{64 \ln \left(v+3 \sqrt{2} t-u\right) v^6}{\left(-9 v-3
   \sqrt{2} t+u\right)^2}-\frac{192 \ln (2) v^6}{\left(-9 v-3 \sqrt{2} t+u\right)^2} \notag \\ &-\frac{32 \ln {v} v^5}{9 v+3
   \sqrt{2} t-u}-\frac{32 \ln \left(v+3 \sqrt{2} t-u\right) v^5}{9 v+3 \sqrt{2} t-u}-\frac{8 v^5}{9 v+3 \sqrt{2}
   t-u}+\frac{96 \ln (2) v^5}{9 v+3 \sqrt{2} t-u} \notag \\ &- \frac{5}{2} \ln (1-v) v^4-\frac{128 \ln {v} v^4}{\left(-9 v-3
   \sqrt{2} t+u\right)^2}+\frac{143}{16} \ln {v} v^4-\frac{5}{2} \ln (v+1) v^4 \notag \\ &-\frac{128 \ln \left(v+3 \sqrt{2}
   t-u\right) v^4}{\left(-9 v-3 \sqrt{2} t+u\right)^2}+\frac{49}{16} \ln \left(v+3 \sqrt{2} t-u\right) v^4-\frac{7}{2}
   \ln \left(9 v+3 \sqrt{2} t-u\right) v^4 \notag \\ &+ 2\ln (u-3 v) v^4-\frac{1}{8} \ln (u-v) v^4+\frac{1}{2} \ln \left(-9 v+3
   \sqrt{2} t+u\right) v^4+\frac{384 \ln (2) v^4}{\left(-9 v-3 \sqrt{2} t+u\right)^2}\notag \\ &+ \frac{65}{16} \ln (2)
   v^4-\frac{33 v^4}{8}+\frac{9 t v^3}{4 \sqrt{2}}+\frac{5 u v^3}{8} -\frac{21 t \ln {v} v^3}{16 \sqrt{2}}+\frac{1}{16}
   u \ln {v} v^3+\frac{64 \ln {v} v^3}{9 v+3 \sqrt{2} t-u} \notag \\ &-\frac{21 t \ln \left(v+3 \sqrt{2} t-u\right) v^3}{16
   \sqrt{2}} -\frac{1}{16} u \ln \left(v+3 \sqrt{2} t-u\right) v^3+\frac{64 \ln \left(v+3 \sqrt{2} t-u\right) v^3}{9 v+3
   \sqrt{2} t-u} \notag \\ &+ \frac{1}{8} u \ln (u-v) v^3 + \frac{16 v^3}{9 v+3 \sqrt{2} t-u}+\frac{63 t \ln (2) v^3}{16
   \sqrt{2}}-\frac{1}{16} u \ln (2) v^3-\frac{192 \ln (2) v^3}{9 v+3 \sqrt{2} t-u} \notag \\ &+ \frac{15}{2} \ln (1-v) v^2 -\frac{3 t
   u \ln {v} v^2}{16 \sqrt{2}}+\frac{64 \ln {v} v^2}{\left(-9 v-3 \sqrt{2} t+u\right)^2}-\frac{71}{4} \ln {v}
   v^2+\frac{15}{2} \ln (v+1) v^2\notag \\ &-\frac{3 t u \ln \left(v+3 \sqrt{2} t-u\right) v^2}{16 \sqrt{2}}+\frac{64 \ln
   \left(v+3 \sqrt{2} t-u\right) v^2}{\left(-9 v-3 \sqrt{2} t+u\right)^2}-\frac{17}{4} \ln \left(v+3 \sqrt{2} t-u\right)
   v^2 \notag \\ &+ \frac{9}{2} \ln \left(9 v+3 \sqrt{2} t-u\right) v^2-8 \ln (u-3 v) v^2- \ln (-v+u-2) v^2-\frac{1}{2}
   \ln (u-v) v^2\notag \\ &- \ln (-v+u+2) v^2+\frac{3}{2} \ln \left(-9 v+3 \sqrt{2} t+u\right) v^2+\frac{9 t u \ln
   (2) v^2}{16 \sqrt{2}}-\frac{192 \ln (2) v^2}{\left(-9 v-3 \sqrt{2} t+u\right)^2}\notag \\ &-\frac{25}{4} \ln (2) v^2+9
   v^2-\frac{6 t v}{\sqrt{2}}-u v+10 \ln (1-v) v+\frac{3 t \ln {v} v}{ \sqrt{2}}+\frac{1}{2} u \ln {v}
   v \notag \\ &-\frac{32 \ln {v} v}{9 v+3 \sqrt{2} t-u}-4 \ln (v+1) v-2\ln \left((v+1)^3\right) v+\frac{3 t \ln \left(v+3
   \sqrt{2} t-u\right) v}{ \sqrt{2}}\notag \\ &-\frac{1}{2} u \ln \left(v+3 \sqrt{2} t-u\right) v-\frac{32 \ln \left(v+3 \sqrt{2}
   t-u\right) v}{9 v+3 \sqrt{2} t-u}-2\ln (-v+u-2) v+ u \ln (u-v) v \notag \\ &+2\ln (-v+u+2) v-2\ln \left(-v^2+u v+3
   \sqrt{2} t-8\right) v+2\ln \left(u^2-v u+3 \sqrt{2} t\right) v\notag \\&-\frac{8 v}{9 v+3 \sqrt{2} t-u}-\frac{9 t \ln (2) v}{
   \sqrt{2}}-\frac{1}{2} u \ln (2 ) v+\frac{96 \ln (2) v}{9 v+3 \sqrt{2} t-u}\notag \\&+4 \ln ^2(1-v)+2\ln ^2{v}+4 \ln
   ^2(v+1)-2\ln ^2(u-v)+2\ln ^2\left(v^2-u v+2\right)\notag \\&+ 8\ln (2) \ln (1-v)-3 \ln (1-v)-4 \ln (1-v) \ln {v}-16
   \ln (2) \ln {v}+6 \ln {v}\notag \\&-8 \ln (1-v) \ln (v+1)-4 \ln {v} \ln (v+1)+8\ln (2) \ln (v+1)-3 \ln
   (v+1)\notag \\&+4 \ln {v} \ln \left(v+3 \sqrt{2} t-u\right)-12 \ln (2) \ln \left(v+3 \sqrt{2} t-u\right)+3 \ln
   \left(v+3 \sqrt{2} t-u\right)\notag \\&+4 \ln {v} \ln \left(v+3 \sqrt{2} t-u+8\right)+4 \ln \left(v+3 \sqrt{2} t-u\right)
   \ln \left(v+3 \sqrt{2} t-u+8\right)\notag \\&-12 \ln (2) \ln \left(v+3 \sqrt{2} t-u+8\right)-4 \ln {v} \ln \left(9 v+3
   \sqrt{2} t-u\right)\notag \\&-4 \ln \left(v+3 \sqrt{2} t-u\right) \ln \left(9 v+3 \sqrt{2} t-u\right)+12 \ln (2) \ln \left(9
   v+3 \sqrt{2} t-u\right)\notag \\&-3 \ln \left(9 v+3 \sqrt{2} t-u\right)-4 \ln (1-v) \ln (-v+u-2)+3 \ln
   (-v+u-2)\notag \\&-4 \ln {v} \ln (u-v)+8\ln (2) \ln (u-v)-4 \ln (v+1) \ln (-v+u+2)\notag \\&+3\ln (-v+u+2)+4 \ln {v}
   \ln \left(-v-3 \sqrt{2} t+u+8\right)\notag \\&+4 \ln \left(v+3 \sqrt{2} t-u\right) \ln \left(-v-3 \sqrt{2} t+u+8\right)-12
   \ln (2) \ln \left(-v-3 \sqrt{2} t+u+8\right)\notag \\&-4\ln (2) \ln \left(v^2-u v+2\right)-3 \ln \left(v^2-u
   v+2\right)-4 \text{Li}_2\left(\frac{1}{2}-\frac{v}{2}\right)-2\text{Li}_2\left(v^2\right)\notag \\&-4
   \text{Li}_2\left(\frac{v+1}{2}\right)+4 \text{Li}_2\left(\frac{v (v-u+2)}{2 (v-1)}\right)-4
   \text{Li}_2\left(-\frac{v+3 \sqrt{2} t-u}{8 v}\right)\notag \\&+2\text{Li}_2\left(\frac{1}{64} \left(v+3 \sqrt{2}
   t-u\right)^2\right)+4 \text{Li}_2\left(\frac{v (-v+u+2)}{2 (v+1)}\right)\notag \\&+26 \ln ^2(2)+3 \ln (2)-\frac{\pi
   ^2}{3}+\frac{5}{2}
\end{align}
where $u=\sqrt{8+v^{2}}$ and  $t = \sqrt{4 + v^2 - v u}$. Our result agrees with the corresponding result in Ref.~\cite{Brown:1991hx} provided that the last term in round bracket in the eleventh line of Eq.~(7) therein
 is $-7y_T^2/4$ instead of  $-7y_T/4$.

\bibliography{biblio}

\end{document}